\documentclass{emulateapj}
\usepackage{apjfonts}

\newcommand {\bband} {B_{435} } 
\newcommand {\vband} {V_{606} } 
\newcommand {\iband} {i_{775} } 
\newcommand {\zband} {z_{850} } 
\newcommand {\solarmassperyear}  {{\rm M_{\sun}~yr^{-1}}}

\shorttitle{DUST OBSCURATION IN LBGs AT $\MakeLowercase{z \sim 4}$}
\shortauthors{HO ET AL.}

\begin{document}

\title{Dust Obscuration in Lyman Break Galaxies at $\MakeLowercase{z \sim 4}$}

\author{I-Ting Ho\altaffilmark{1}, Wei-Hao Wang\altaffilmark{1}, Glenn E. Morrison\altaffilmark{2,3},
and Neal A. Miller\altaffilmark{4}}
\altaffiltext{1}{Institute of Astronomy \& Astrophysics, Academia Sinica, P.O. Box 23-141, Taipei 10617, Taiwan}
\altaffiltext{2}{Institute for Astronomy, University of Hawaii, Honolulu, HI 96822, USA}
\altaffiltext{3}{Canada-France-Hawaii Telescope, Kamuela, HI 96743, USA}
\altaffiltext{4}{Department of Astronomy, University of Maryland, College Park, MD 20742, USA}

\email{itho@ifa.hawaii.edu; itho@asiaa.sinica.edu.tw}
\slugcomment{ApJ accepted, August 2010}

\begin{abstract}
Measuring star formation rates (SFRs) in high-$z$ galaxies with their rest-frame ultraviolet (UV) continuum can be uncertain because of dust obscuration. Prior studies had used the submillimeter emission at $850~\mu\rm m$ to determine the intrinsic SFRs of rest-frame UV selected galaxies, but the results suffered from the low sensitivity and poor resolution ($\sim15\arcsec$).  Here, we use ultradeep Very Large Array 1.4 GHz images with $\sim1\arcsec$--$2\arcsec$ resolutions to measure the intrinsic SFRs. We perform stacking analyses in the radio images centered on $\sim3500$  Lyman break galaxies (LBGs) at $z\sim4$ in the Great Observatories Origins Deep Survey-North and South fields selected with {\it Hubble Space Telescope}/Advanced Camera for Surveys data. The stacked radio flux is very low, $0.08\pm0.15$ $\mu$Jy, implying a mean SFR of $6\pm11~\solarmassperyear$.  This is comparable to the uncorrected mean UV SFRs of $\sim5~\solarmassperyear$, implying that the $z\sim4$ LBGs have little dust extinction.
The low SFR and dust extinction support the previous results that $z\sim4$ LBGs are in general not submillimeter galaxies. We further show that there is no statistically significant excess of dust-hidden star-forming components within $\sim22$~kpc from the LBGs. 

\end{abstract}


\section{INTRODUCTION}{\label{introduction}}

Understanding galaxy evolution in the early Universe requires large samples of various kinds of high-$z$ galaxies.  One important technique enabling selections of $z>2.5$ star-forming galaxies is the Lyman break technique (\citealp{cow88,son90,lil91,ste93,ste95}; also see a review in \citealp{gia02}). The UltraViolet (UV) spectrum of a high-redshift star-forming galaxy usually exhibits a Lyman continuum discontinuity at 912 \AA, which is caused by absorption of \ion{H}{1} in the stellar atmosphere of massive stars, and interstellar and intergalactic media. By using this feature, large samples of Lyman break galaxies (LBGs) can be selected in various redshift ranges by searching for sudden brightness dropouts between two adjacent broad-band images.

An important property of LBGs is their star formation rates (SFRs). However, determining SFRs for $z>3$ systems is still challenging because most of the diagnostics at low redshifts are unavailable when light becomes dimmer and redshifts to wavelengths that are harder to observe. 
One of the most commonly used method to determine SFRs at high redshift is to measure the rest-frame UV continuum which shifts at $z>3$ to the optical and near-infrared \citep[e.g.,][]{ken98}. Unfortunately, the UV continuum can be easily attenuated by dust, which results in underestimation of the SFRs. \citet{bou09}  attempted to determine the correction factor by using the observed correlation between the ratio of far-infrared (FIR) to UV flux and the UV spectral slope \citep{meu99}. They found that UV continuum of $z\sim4$ LBGs in the Great Observatories Origins Deep Survey-North (GOODS-N) and South (GOODS-S) fields underestimates SFRs by factors of $\sim3$--6. However, this result is debatable since the correlation exhibits significant scatter in different populations of galaxies \citep[e.g.,][]{cor06,how10}.

A more direct way to estimate the SFRs is to deduce them at longer wavelengths where light is not affected by dust obscuration. Efforts had been made using the submillimeter $850~\mu\rm m$ emission to determine the intrinsic SFRs in LBGs \citep{pea00,cha00,web03}. However, these results have low signal-to-noise ratios (S/Ns), presumably because the submillimeter single-dish maps were not deep enough, a consequence of instrumental, sky, and confusion noises (i.e., uncertainties contributed by nearby bright sources or faint undetected sources). In addition, an assumption of dust temperature is required to estimate the total infrared (IR) luminosity. These all make the estimate of SFRs based on submillimeter measurements quite uncertain. 

The radio wavelength is an alternative probe of SFRs. At 1.4 GHz, the radio continuum is dominated by synchrotron radiation from relativistic electrons produced by supernovae. By converting the 1.4 GHz luminosity to total IR luminosity with the well-known radio---FIR correlation \citep{con92}, it is possible to estimate intrinsic SFRs \citep{ken98}. The radio---FIR correlation is fairly insensitive to dust temperature. The angular resolution in the radio can be very high ($\sim1\arcsec$--$2\arcsec$), meaning very little confusion noise. Because of these great advantages, this radio-based method has been used in various high-{\it z} studies \citep[e.g.,][]{red04,wang06,car08,pan09}.

The Very Large Array (VLA) provides high resolution in the radio, but its sensitivity is insufficient for directly detecting normal star-forming galaxies at $z\gtrsim1.5$. Therefore, stacking radio fluxes of large samples of LBGs to determine their mean SFRs is a necessary approach.  \citet{red04} stacked $z\sim2$ LBGs and found the dust correction to be $\sim4.5$. Using a similar method, \citet{car08} determined the dust correction for $z\sim3$ LBGs to be 1.8.

In order to determine the intrinsic SFRs for $z\sim4$ LBGs, we performed radio stacking analyses in the GOODS-N and GOODS-S fields. The GOODS {\it Hubble Space Telescope/\rm Advance Camera for Surveys} ({\it HST/\rm ACS}) data \citep{gia04} were used to locate $\sim3500$ LBGs in the fields, and deep VLA 1.4 GHz images \citep{mil08,mor10} were used to measure and stack the radio fluxes. 
We note that although radio stacking of $z\sim4$ LBGs had also been attempted by \citet{car08}, we used much deeper optical images ($\sim2$ mag deeper at {\it V} band ) and radio images ($\sim15\%-40\%$ deeper). We are therefore probing much closer to the typical members in the $z\sim4$ LBG population, as in \citet{bou09}.

In this paper, we first describe the observational data in Section~\ref{data}. In Section~\ref{method}, we describe our methods of selecting LBGs, stacking analysis, and estimations of  SFRs using radio and UV. Finally, we compare the two different SFRs and discuss the implications in Section~\ref{discussion}, and give a summary in Section~\ref{summary}. We assume $\Omega_0=0.3$, $\Omega_\Lambda=0.7$, and $H_0=70~\rm km~s^{-1}~Mpc^{-1}$. All magnitudes are in the AB magnitude system.

\section{DATA}\label{data}

\subsection{\it HST/\rm ACS}\label{hst}
The {\it HST/\rm ACS} multiband imaging in the GOODS fields \citep{gia04} consists of four passbands: F435W, F606W, F775W, and F850LP, which are referred to as $\bband$, $\vband$, $\iband$, and $\zband$, respectively. 
We adopt the v2.0 source catalogs for selecting our LBG samples. Quantities measured with SExtractor \citep{ber96} ``automatic aperture'' ({\tt MAG\_AUTO}, {\tt FLUX\_AUTO}, etc.) are used to approximate total values. 

\subsection{VLA}\label{vla}

The VLA 1.4 GHz images of the GOODS-N and GOODS-S fields are described in more details in \citet{mor10} and \citet{mil08}, respectively. In brief, the North field has an rms noise of $\sim4~\mu\rm Jy~beam^{-1}$ at the center, and $<6~\mu\rm Jy~beam^{-1}$ at the edges of the ACS fields. The South field has an rms noise of $\sim6~\mu\rm Jy~beam^{-1}$ at the center, and $<7.5~\mu\rm Jy~beam^{-1}$ at the edges of the fields. These sensitivities have been corrected for the primary beam response. The beam FWHM of the North field is $1\farcs7 \times1\farcs6$, which corresponds to $\rm11.8~kpc\times11.1~kpc$ at $z=4$. The beam FWHM of the South field is $2\farcs8\times1\farcs6$, which corresponds to $\rm19.5~kpc\times11.1~kpc$ at $z=4$. Data Release 2 of the South field is used here.

\section{METHOD}\label{method}

\subsection{Sample Selection}\label{sample selection}

\begin{figure}[h!]
\plotone{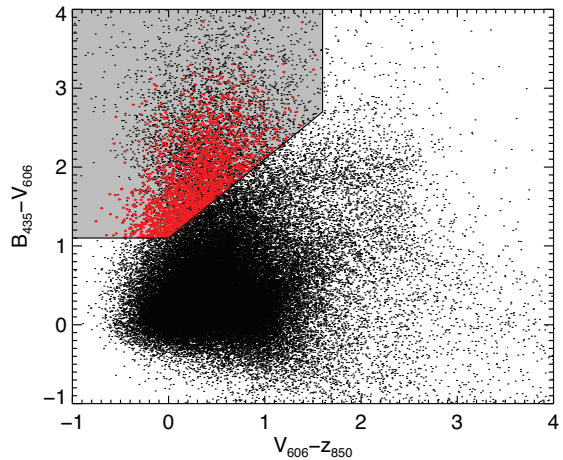}
\caption{A color-color diagram showing the $B$-dropouts selection criteria. Objects in the GOODS fields are plotted with dots. 
Intersection of the three color selection criteria described in Section~\ref{sample selection} are shaded in grey. The selected LBGs are shown with red dots. Note that some of the objects inside the grey region are not selected as LBGs because they do not meet the other criteria (e.g., low S/N or too compact).
\label{figure1}}
\end{figure}

``{\it B}-dropout'' galaxies at $z\sim4$ are selected utilizing the redshifted Lyman break located between $\bband$ and $\vband$. We adopt the well established criteria \citep[e.g.,][]{bec06,bou07}:
\[\bband -  \vband  >1.1,\\\]
\[\bband-\vband > (\vband-\zband)+1.1,\\\]
\[\vband-\zband <1.6,\\\]
\[S/N(\vband)>5, \ \ \  {\rm and} \ \ S/N(\iband)>3.\]
In addition, compact objects (SExtractor stellarity indices greater than 0.8) with $\iband<$ 26.5 are rejected from our sample to prevent stellar contamination. 
A color-color diagram illustrating the sample selection is shown in Figure~\ref{figure1}. 
In total, we selected 1778 and 1679 {\it B}-dropouts in the North and South fields, respectively. These selection criteria efficiently select LBGs between $z\sim3$ and $z\sim4.5$, with a mean redshift of 3.8 \citep[see more details in][]{bou07}. The number of LBGs selected by us is very similar to that in \citet{bou07}, who used similar GOODS ACS data. 

\subsection{Stacking Analyses}\label{stacking analyses}
We measured the radio fluxes of the $z\sim4$ LBGs with aperture photometry at their optical positions. Small offsets ($<0''.3$) between the coordinates of the radio and optical images had been corrected utilizing their source catalogs. Apertures of two different sizes were used here: small-apertures with radii equal to the beam FWHMs  ($\sim11$ kpc at $z=4$); large-apertures with radii equal to twice the beam FWHMs ($\sim22$ kpc at $z=4$). Here we first focus on the small aperture results and we will discuss the large aperture results in Section~\ref{implications}. Elliptical apertures were used for the South field to match its beam shape. Fluxes measured with our aperture photometry were calibrated with the radio catalogs \citep{mil08,mor10}. For sources in the radio catalogs, we measured their fluxes with our aperture photometry and computed the median of the ratios between our fluxes and the catalog fluxes. The ratios, which were found to be very close to unity for the aperture sizes we used, were applied back to our radio fluxes.

Sources with radio fluxes $>100~\mu \rm Jy$ were rejected from our measurements because their radio fluxes do not likely reflect their true SFRs.  There are eight such sources. If they are normal star-forming galaxies at $z=4$, their IR luminosity would be $\gtrsim5\times 10^{13}~L_\sun$ (or SFR $\gtrsim7000~\solarmassperyear$), which is unusually large. We inspected these sources individually. They either are affected by nearby bright radio sources that do not appear to be at the same redshifts of the LBGs, or do not appear to be associated with known submillimeter galaxies (SMGs) in the samples of \citet{wang04}, \citet{per08}, \citet{dev09}, and \citet{wei09}. Therefore, they are highly unlikely high-redshift ultraluminous starbursting galaxies. To avoid the bias from unrelated nearby bright sources and from active galactic nuclei (AGNs), we did not include these eight sources in our stacking analyses.   For those sources with radio fluxes $<100~\mu\rm Jy$, we then averaged their radio fluxes and subtracted a background value (see below) from the means to get the final stacked radio fluxes.

The uncertainties of our stacked fluxes were estimated with Monte Carlo simulations, with an assumption that the galaxies are distributed randomly over the map. We measured the mean radio fluxes at random positions, and the number of random positions is the same as that of the $<100~\mu \rm Jy$ {\it B}-dropouts. The same apertures were used and the same rejection criterion of $<100~\mu \rm Jy$ was adopted. This measurement is repeated 10,000 times, and the mean radio fluxes have a fairly Gaussian-like distribution. We then calculated the mean and the dispersion of these 10,000 measurements. The mean was subtracted from the stacked radio flux of the LBG sample to form the final stacked radio flux. This subtraction is to account for the effect of imperfect clean and chance projection of random radio sources in our flux apertures. Likewise, the dispersion can then represent the uncertainty of the stacked radio flux.

There are two caveats in the above procedures. First, if $z\sim4$ LBGs are clustered at scales similar to the sizes of our apertures, our assumption that LBGs are randomly distributed would break down. Our stacking method would then overestimate the radio fluxes.  We investigated clustering with the same method used by \citet{mar09}.  We found that the variance-to-mean ratio of source numbers inside randomly placed $r<10\arcsec$ circles of our sample exceeds that of random distribution by less than 0.1, and becomes smaller with smaller radii. Therefore, we conclude that there is no significant small-scale clustering in our LBG sample and the flux overestimate caused by small-scale clustering can be neglected.
 
Second, the faulty estimate of uncertainties can be caused by concentration of sources in high or low sensitivity regions (i.e., the slightly uneven sensitivity distribution caused by the primary beam falloff in the GOODS-N or by the mosaicking in the GOODS-S). Instead of stacking at random positions to estimate the uncertainty, we also carried out the simulation based on the source positions of the real LBG sample. We added an offset to the positions of the LBGs and stacked their fluxes.  We repeated this for 10,000 times with different random offsets of $15\arcsec-60\arcsec$, and calculated the mean and the dispersion. The results are within 30\% to those measured from random positions, suggesting that LBGs are not concentrated in specific regions.

With the above methods, we measured the final stacked fluxes and flux errors of the LBGs in the two fields separately. For each aperture size, we combined the two final stacked fluxes by weighting them with the inverse of the square-errors to form the combined flux. The results are summarized in Table~\ref{tablestacking}.  The stacked radio signal is very weak, consistent with zero within the noise.

\subsection{Star Formation Rate}

\subsubsection{Radio SFR}{\label{radio sfr}}
The stacked radio fluxes can be converted to SFRs. Assuming a universal synchrotron emission spectral index of $\alpha=-0.8$, we convert the stacked radio fluxes $S_{\rm stack}$ to the rest-frame 1.4 GHz luminosity densities $L_{\rm1.4GHz}$, i.e., 
\begin{equation}
L_{\rm1.4GHz}=4\pi d_l^2\sl S_{\rm stack} \sl(1+z)^{-(1+\alpha)},
\end{equation}
where $d_l$ is the luminosity distance and $z$ is the mean redshift of 3.8. 
With the local radio---FIR correlation \citep{hel85,con92}, $L_{\rm1.4GHz}$ can be converted to FIR luminosity $L_{\rm FIR(40-120)}$, approximately the total luminosity between $\lambda=42.5~\mu \rm m$ and $\lambda=122.5~\mu \rm m$, with the following relation:
\begin{equation}
q=log{{L_{\rm FIR(40-120)}}\over{3.75\times10^{12}\rm~W}}\sl - log{{{L_{\rm1.4GHz}}}\over{\rm W~Hz^{-1}}},
\end{equation}
with $q=2.34\pm0.01$\citep{yun01}. Although this value was derived locally ($z\lesssim0.15$), recent studies suggest weak evolution out to $z\sim2$ \citep{sar10,ivi10a,ivi10b}. Extrapolating $q$ to $z=4$ using the relation $q\propto(1+z)^{\gamma}$ with $\gamma=-0.04\pm0.03$ \citep{ivi10b} implies that $L_{\rm FIR}$ (and therefore radio SFRs) will be lowered by only $\sim30\%\pm20\%$, which is insignificant compared to the uncertainties in our results.

SFRs can then be estimated with the conversion in \citet{ken98}: 
\begin{equation}
{{\rm SFR\over (\solarmassperyear)}}=4.5\times10^{-44} { L_{ \rm IR(8-1000)} \over \rm erg~s^{-1}},
\end{equation}
where $L_{\rm IR(8-1000)}$ is approximately the total luminosity between $8~\mu\rm m$ and $1000~\mu\rm m$. This estimation is within $\sim30\%$ to other published calibrations. 
We derived the ratio between $L_{\rm FIR(40-120)}$ and $L_{\rm IR(8-1000)}$ by computing these two quantities on model spectral energy distributions of six nearby normal and starburst galaxies \citep{sil98}, using their original definitions over the {\it Infrared Astronomical Satellite}{\it (IRAS)} bands \citep[see the summary in][]{san96}. 
The $L_{\rm IR(8-1000)}$ to $L_{\rm FIR(40-120)}$ ratio has a range of $1.71-2.32$, with a weak anti-correlation between the ratio and IR luminosity and  a mean of 2.05. This mean ratio is slightly higher than that used in \citet{yun01} of 1.5, which is based on measurements in luminous {\sl IRAS} galaxies and starburst galaxies \citep{san91,meu99,cal00}, but is closer to the values in nearby normal spiral galaxies and low luminosity starbursts ($\sim2$ for M~51 and M100, and 2.3 for M~82, based on the Silva et al.\ templates). Using this value, we derived the SFRs listed in Table~\ref{tablesfr}, which are consistent with normal galaxies or low-luminosity starbursts.

\begin{table}
\begin{center}
\caption{Mean Radio Fluxes \label{tablestacking}}
\begin{tabular}{cccccc}
\hline\hline
Field& \multicolumn{2}{c}{Small Aperture\tablenotemark{a}} &\multicolumn{2}{c}{Large Aperture\tablenotemark{b}}\\
& $F_{\rm LBG}$  & $F_{\rm random}$ & $F_{\rm LBG}$  & $F_{\rm random}$ \\
&($\mu\rm Jy$)  & ($\mu\rm Jy$)& ($\mu\rm Jy$)  & ($\mu\rm Jy$) \\
\hline
GOODS-N&$-0.05\pm0.18$&$0.36\pm0.18$&$0.29\pm0.28$&$1.10\pm0.28$\\
GOODS-S&$0.36\pm0.26$&$0.10\pm0.26$&$0.54\pm0.54$&$-0.03\pm0.54$\\
\hline
Combined&$0.08\pm0.15$&-&$0.34\pm0.25$&-\\
\hline
\end{tabular}
\tablecomments{$F_{\rm random}$ is the mean radio flux of random positions derived
with the Monte Carlo simulation in \protect Section~\ref{stacking analyses}. This mean is subtracted from the
measured flux of the LBG sample, and its error is propagated to the mean of the LBG sample. $F_{\rm LBG}$ listed has already been 
subtracted by $F_{\rm random}$. }
\tablenotetext{1}{Apertures with radii equal to the beam FWHM.}
\tablenotetext{2}{Apertures with radii equal to twice the beam FWHM.}
\end{center}
\end{table}

\subsubsection{UV SFR}
The $\iband$ and $\vband$ bands correspond to rest-frame UV, and can be used to calculate SFRs, {\it uncorrected for extinction} \citep{mau98}, with 
\begin{equation}
{\rm SFR}(\solarmassperyear)={ \sl L_{\rm UV}({\rm ergs~s^{-1}~Hz^{-1}}) \over  C},
\end{equation} 
where $L_{\rm UV}$ is the UV luminosity density at the same redshift of 3.8, and $C$ is $8.0\times10^{27}$ and $7.9\times10^{27}$ at 1500\AA\ and 2800\AA, respectively. We list in Table~\ref{tablesfr} the mean UV SFRs of the {\it B}-dropouts with radio fluxes $< 100~\mu \rm Jy$. The uncorrected SFRs derived from $\iband$ and $\vband$ are remarkably similar to each other as well as to that derived from the radio stacking analyses, indicating very little extinction in the observed rest-frame UV emission.

\begin{table}
\begin{center}
\caption{Star Formation Rates Derived with Various Fluxes\label{tablesfr}}
\begin{tabular}{cccccc}
\hline\hline
&\multicolumn{2}{c}{Radio}&\multicolumn{2}{c}{Optical}\\
&Small aperture&Large aperture&$\iband$&$\zband$\\
\hline
SFRs($\solarmassperyear$)&$6.0\pm11.0$&$25.7\pm18.8$&4.92&5.14\\
\hline
\end{tabular}
\end{center}
\end{table}

\section{RESULT AND DISCUSSION}\label{discussion}

\subsection{Dust extinction in $z\sim4$ LBGs}
\begin{figure}[h!]
\plotone{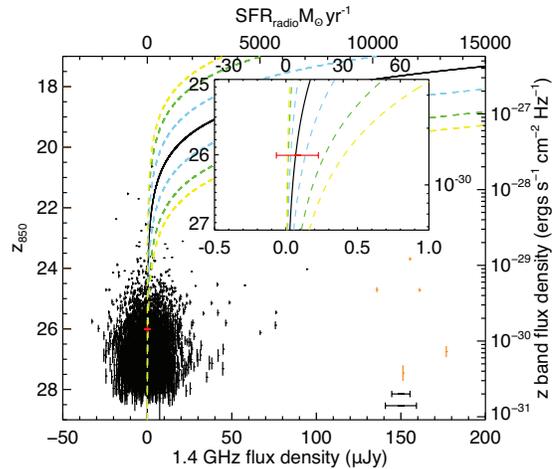}
\caption{{\it HST} $\zband$ magnitude vs. 1.4 GHz {\it VLA} flux of the $\sim3500$ {\it B}-dropouts in the GOODS fields. Orange symbols indicate objects not included in our stacking analyses. The radio fluxes are measured with small apertures with radii equal to the beam FWHMs ($\sim11~\rm kpc$ at $z=4$). The $\zband$ uncertainty of each data point is indicated with vertical bars. The radio uncertainties, which depend on the locations on the maps, are not shown to avoid confusion. Instead, we plot the typical radio flux uncertainties for galaxies close to the center of the 2 fields at the lower-right corner for reference (upper: North, lower: South). 
The inferred SFR corresponding to the measured flux density at 1.4 GHz (see text for details) is shown on the upper x-axis. 
The average of all the {\it black} data points are shown with the red symbols, with the radio uncertainties estimated from the MonteCarlo simulation. The black curves indicate the locus of equal SFR deduced from the $\zband$ and the radio. The color coded curves, from left to right, indicate 1/6, 1/4, 1/2, 2, 4, and 6 times dust corrections from the UV to radio SFR. A blow up to the stacked data point is shown as an insert. \label{figure2}}
\end{figure}

In Figure~\ref{figure2}, we plot the $\zband$ magnitudes versus the small-aperture radio fluxes and the corresponding SFRs of all the {\it B}-dropouts in the two fields, with the red points indicating the mean ($<100~\mu\rm Jy$ sources). As can be seen in the figure and Table~\ref{tablesfr}, the {\it average} SFRs deduced from the radio and the rest-frame UV are consistent within measurement uncertainties. The relatively small SFR derived from the radio flux suggests little extinction in the $z\sim4$ LBGs. Unfortunately, although this is by far the deepest radio stacking on $z\sim4$ LBGs ($\sim3\times$ deeper than that in \citealp{car08}), the error in the stacked radio flux is still too large to pin down the extinction correction. An extinction correction of $>6$ is ruled out at 2.2 $\sigma$ and an extinction correction of $>4$ is ruled out at 1.3 $\sigma$. \citet{bou09} derived extinction corrections of $\sim3-6$ for $z\sim4$ LBGs based on their UV continuum slopes. Our result slightly favors a lower value and is marginally ($\sim2~\sigma$) consistent with the result of \citet{bou09}. 

We note that an even stricter upper limit on the radio SFR can be placed if any of the following is true: (1) the radio---FIR correlation $q$ factor exhibits a decline with redshift (see Section~\ref{radiofir}); or (2) the conversion from $L_{\rm FIR(40-120)}$ to $L_{\rm IR(8-1000)}$ is overestimated (see Section~\ref{radio sfr}); or (3) there is AGN contribution in our radio SFR (see Section~\ref{agn}). The combined effect of all the above may not be negligible and can make the extinction correction for $z\sim4$ LBGs even lower.  However, this remains to be tested by future observations.

\citet{car08} found a low extinction correction of 1.8 on $z\sim3$ LBGs.  On the other hand, on $z\sim4$ LBGs, they had a 2~$\sigma$ detection of $0.83\pm0.42$ $\mu$Jy on 1447 $B$-dropouts, which also suggests a low extinction correction.  This radio flux is much higher than ours ($0.08\pm0.15$ $\mu$Jy) but their ground-based $V$-band limiting magnitude is also much shallower than ours (by $\sim2$ mag). In other words, we are probing different LBG luminosities. Comparing our $z\sim4$ result with the $z\sim4$ result in \citet{car08} is therefore not straightforward.


\subsection{AGN contamination}\label{agn}
To avoid AGN contamination, a flux cut of $<100~\mu\rm Jy$ was set and eight sources were rejected (one in the north and seven in the south). If these sources comprise {\it all} the AGNs in our sample, our sample would have a very low AGN fraction of $\sim0.2\%$. For comparison, the AGN fraction in LBGs at $z\sim3$ was estimated to be $\sim3\%$, based on optical spectroscopy \citep{sha03}. This difference could be due to the fact that our LBGs are much less luminous than those in the spectroscopic sample. 

It is possible to roughly separate AGNs and star forming galaxies based on the radio power. In the local radio luminosity function \citep[e.g.,][]{mau07}, the 1.4 GHz power that divides AGNs and starbursts is $\sim10^{23}~\rm W~Hz^{-1}$. \citet{cow04} show that this AGN/starburst division increases to $\sim10^{24}~\rm W~Hz^{-1}$ at $z\sim1$.  This power corresponds to $\sim10$ $\mu$Jy for $z=4$. However, the increasing trend observed by \citet{cow04} can continue to $z\gg1$, moving the AGN/starburst division to a radio flux greater than 10 $\mu$Jy.  Furthermore, high-redshift flat-spectrum radio sources (presumably AGNs) have a more negative $K$-correction than that of steep-spectrum sources (starbursts). This also increases the apparent flux division between AGNs and starbursts at observed-frame 1.4 GHz (by roughly a factor of 2). We thus expect sources below our 100 $\mu$Jy cut to be dominated by star forming galaxies.

It is interesting to test what happens if we assume that $>50$ $\mu$Jy sources are also AGNs and remove them from the stacking analyses. To do this, we rejected 13 sources that are brighter than 50 $\mu$Jy and introduced the same 50 $\mu$Jy cut in the Monte Carlo simulation.  We found a stacked flux of $0.05\pm0.14$ $\mu$Jy, corresponding to a mean SFR of $3.8\pm10.5$ $\solarmassperyear$.  The stacked flux decreases further if we keep lowering the flux cut but the change is well within the noise.  Nevertheless, it is obvious that we will get a lower mean SFR if we assume more AGN contribution.  Although we are limited by noise here, our conclusions that $z\sim4$ LBGs have low SFRs and the mean extinction correction seems to be lower than 5 are thus still valid.

\subsection{The Radio---FIR Correlation}\label{radiofir}

\citet{car08} discussed the use of the local radio---FIR correlation to derive SFRs of high-redshift galaxies. They mentioned several possibilities that this correlation may be different at high redshift. Indeed, a weak evolution has been suggested by recent studies \citep{sar10,ivi10a,ivi10b}. However, as mentioned in Section~\ref{radio sfr}, even if we extrapolate the evolution based on a declining $q$ \citep{ivi10b} and adopt a $L_{\rm IR(8-1000)}$ to $L_{\rm FIR(40-120)}$ ratio that is 25\% lower, our main conclusions would not change and this would make the SFR of the LBGs even lower. Furthermore, the local radio---FIR correlation is only used to calibrate the relation between SFR and radio power.  Even if there is an evolution in the radio---FIR correlation at high redshift, we cannot be certain that the evolution arises from the radio part of the correlation and affects the radio---SFR relation.  In short, we see neither evidence nor strong argument that the radio SFR is significantly biased by the assumption of the local radio--FIR correlation.

\subsection{Implications for SMGs}\label{implications}

The relatively low dust correction and SFRs imply that $z\sim4$ LBGs in general do not have extreme star formation activities enshrouded in dust, unlike that in SMGs. If the total SFR of our $z\sim4$ LBGs is only contributed by SMGs with SFR of $1000~\solarmassperyear$ and if other non-SMG LBGs have no star forming activities, then there can be no more than $21\pm39$ SMGs in our LBG sample. 
In comparison, the source counts at $850~\mu\rm m$ \citep{cop06} suggest $\sim350$ SMGs in a field of the combined GOODS size, with $S_{850}\rm >1.4~mJy$ and a mean SFR of $1000~\solarmassperyear$. This means that even if the total SFR of our LBG sample is dominated by $z\sim4$ SMGs, they can only account for $<10\%$ of the entire SMG population at $S_{850}\rm>1.4~mJy$. This implies that either LBGs and SMGs have little overlap at $z\sim4$ or most SMGs are at $z\ll4$. We note that previous studies in the submillimeter showed that LBGs are in general not SMGs \citep{cha00,pea00,web03}, and that most SMGs are at $z<3.5$ \citep{cha03,cha05}. Our results are consistent with these.

\begin{figure}
\plotone{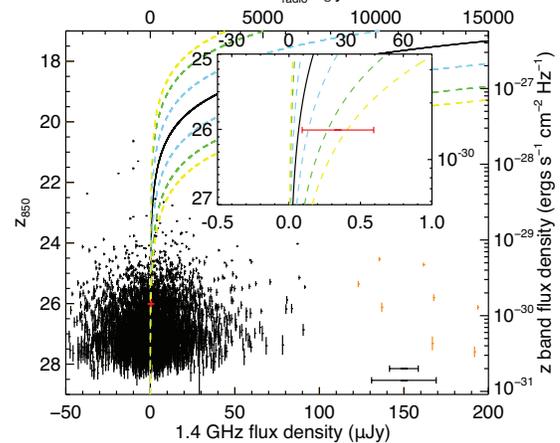}
\caption{Same as Figure~\ref{figure2}, but the radio fluxes are measured with apertures of radii twice the beam FWHM ($\sim22~\rm kpc$ at $z=4$). \label{figure3}}
\end{figure}

Another scenario related to SMGs worth investigating is dust-hidden companions sitting close to the LBGs. An example of this is the $z=4.5$ SMG in \citet{cap08}. This galaxy is independently selected as an LBG, and has an ultraluminous dusty component showing up at $>3.6\mu\rm m$ next to its rest-frame UV emission. To know whether similar systems are common at $z\sim4$, we performed the same radio stacking using the large apertures with radii twice the beam FWHM ($\sim22$ kpc at $z=4$) to measure the radio fluxes. As shown in Figure~\ref{figure3} and Table~\ref{tablesfr}, the radio SFR is slightly higher but still consistent with the UV SFR, implying that there is no statistically significant excess of dust-hidden star-forming components close to LBGs.


\section{Summary}\label{summary}
We employed radio stacking analyses on $\sim3500$ LBGs at $z\sim4$ in the GOODS-N and GOODS-S fields. The mean radio flux was converted to FIR luminosity via the well-known radio---FIR correlation, and then converted to an intrinsic SFR, which is $6\pm11\solarmassperyear$. By comparing with SFRs estimated from rest-frame UV, we found a maximum UV extinction correction of $\sim6$. This is roughly consistent with that derived from UV continuum slopes by \citet{bou09} with a very similar sample. The low extinction correction confirms that $z\sim4$ LBGs are in general not SMGs.  We also investigated the possibility of finding dust-hidden companions close to the LBGs at $z\sim4$. The stacked radio flux within $\sim22\rm~kpc$ to the LBGs implies a mean SFR that is still consistent with the UV inferred SFR, which suggests no statistically significant excess of dust-hidden star-forming components close to LBGs.

\acknowledgments
We thank the referee for providing comments that improve this paper. W.-H.W. and I.-T.H. acknowledge a grant from the National Science Council of Taiwan (98-2112-M-001-003-MY2).

\end{document}